\documentclass{aa}
\usepackage[version=4]{mhchem}
\usepackage{gensymb}
\usepackage[varg]{txfonts}
\usepackage{xcolor}
\usepackage{soul}
\usepackage{natbib,twoopt}
\bibpunct{(}{)}{;}{a}{}{,}             %% natbib format for A&A and ApJ
\sethlcolor{green}

\title{Photolysis-induced scrambling of PAHs as a mechanism for deuterium storage}

\author{Sandra D. Wiersma\inst{\ref{inst1},\ref{inst2}}
\and Alessandra Candian\inst{\ref{inst1},\ref{inst3}}
\and Joost M. Bakker\inst{\ref{inst2}}
\and Jonathan Martens\inst{\ref{inst2}}
\and Giel Berden\inst{\ref{inst2}} 
\and Jos Oomens\inst{\ref{inst1},\ref{inst2}}
\and Wybren Jan Buma\inst{\ref{inst1},\ref{inst2}}
\and Annemieke Petrignani\inst{\ref{inst1}}\thanks{\email{a.petrignani@uva.nl}}
}

\institute{
Van `t Hoff Institute for Molecular Sciences, University of Amsterdam, Science Park 904,
1090 GD, Amsterdam, The Netherlands \label{inst1}
\and
Radboud University, Institute for Molecules and Materials, FELIX Laboratory, Toernooiveld 7, 6525 ED Nijmegen, The Netherlands\label{inst2}
\and
Leiden Observatory, Leiden University, Niels Bohrweg 2, 2300 RA Leiden, The Netherlands\label{inst3}
}

\abstract{
\textit{Aims} We investigate the possible role of polycyclic aromatic hydrocarbons (PAHs) as a sink for deuterium in the interstellar medium (ISM) and study UV photolysis as a potential underlying chemical process in the variations of the deuterium fractionation in the ISM.

\textit{Methods} The UV photo-induced fragmentation of various isotopologs of deuterium-enriched, protonated anthracene and phenanthrene ions (both \ce{C14H10} isomers) was recorded in a Fourier Transform Ion Cyclotron Resonance Mass Spectrometer (FTICR MS). Infrared multiple photon dissociation (IRMPD) spectroscopy using the Free-Electron Laser for Infrared eXperiments (FELIX) was applied to provide IR spectra. Infrared spectra calculated using density functional theory (DFT) were compared to the experimental data to identify the isomers present in the experiment. Transition-state energies and reaction rates were also calculated and related to the experimentally observed fragmentation product abundances.

\textit{Results} The photofragmentation mass spectra for both UV and IRMPD photolysis only show the loss of atomic hydrogen from \ce{[D-C14H10]+}, whereas \ce{[H-C14D10]+} shows a strong preference for the elimination of deuterium. Transition state calculations reveal facile 1,2-H and -D shift reactions, with associated energy barriers lower than the energy supplied by the photo-excitation process. Together with confirmation of the ground-state structures via the IR spectra, we determined that the photolytic processes of the two different PAHs are largely governed by scrambling where the H and the D atoms relocate between different peripheral C atoms. The $\sim$0.1 eV difference in zero-point energy between \ce{C-H} and \ce{C-D} bonds ultimately leads to faster H scrambling than D scrambling, and increased H atom loss compared to D atom loss.
                     
\textit{Conclusion} We conclude that scrambling is common in PAH cations under UV radiation. Upon photoexcitation of deuterium-enriched PAHs, the scrambling results in a higher probability for the aliphatic D atom to migrate to a strongly bound aromatic site, protecting it from elimination. We speculate that this could lead to increased deuteration as a PAH moves towards more exposed interstellar environments. Also, large, compact PAHs with an aliphatic \ce{C-HD} group on solo sites might be responsible for the majority of aliphatic \ce{C-D} stretching bands seen in astronomical spectra. An accurate photochemical model of PAHs that considers deuterium scrambling is needed to study this further.
}


\begin{document}

\maketitle

\section{Introduction}
 Nearly all deuterium in our Universe was formed during the nucleosynthesis era after the Big Bang \citep{Liddle2003}. Since then, it has mostly been depleted through stellar nucleosynthesis leading to variations in its interstellar abundance \citep{Epstein1976,ReevesHubert;AudouzeJean;FowlerWilliamA.;Schramm1973}. The dispersion in deuterium abundance shows strong ties to the so-called metallicity of the environment, that is, the abundance of elements other than H and He. However, not all of the dispersion can be attributed to nucleosynthesis. The missing deuterium is likely chemically stored in molecules and grains, making the local interstellar deuterium abundance a direct tracer of chemical activity \citep{Linsky2006,Draine2006,Roueff2007}. 

A family of molecules in which the missing interstellar deuterium could be stored is that of the polycyclic aromatic hydrocarbons \citep[PAHs;~][]{Peeters2004,Hudgins2004,Draine2006,Onaka2013,Buragohain2015,Doney2016,Buragohain2016}. It has been estimated that about 85\% of the carbon in dust is aromatic \citep{Pendleton2002}, and that PAHs bear about 5 to 10 \% of all cosmic carbon \citep{Tielens2013}. Their large heat capacity and stable aromatic nature allow them to withstand harsh radiative conditions, and survive by re-emitting the absorbed radiation
in well-defined IR spectral regions widely known as the aromatic infrared bands (AIBs). The AIBs are commonly associated with aromatic \ce{C-C} and \ce{C-H} vibrations, and are observed throughout the interstellar medium (ISM) in the 3-18 $\mu$m spectral range \citep{Tielens2008,Peeters2011,Tielens2013}. Bands in this range are observed in different types of interstellar sources, showing PAHs to be present under different conditions \citep{Ricca2011,Tielens2013}. Although the AIBs show that aromatic species exist in space, no individual PAH has been identified to date. The only aromatic molecules that have been firmly identified are the possible PAH precursors benzene \citep{Cernicharo2001,Kraemer2006} and benzonitrile \citep{McGuire2018}, and the fullerenes \ce{C60} and \ce{C70} \citep{Cami2010,Sellgren2010,Berne2013,Campbell2015, Cordiner2019}; the latter two could be formed by the radiative processing of larger PAHs \citep{berne2012}. The ubiquitous presence of interstellar PAHs is further supported by the identification of several PAHs in the Murchison and Allende carbonaceous chondrites. \ce{^13C}/\ce{^12C} isotopic studies confirm the interstellar origin of the PAHs on these chondrites \citep{Kerridge1987,Spencer2008}. 

The presence of deuterated PAH species in the ISM is revealed through the observation of \ce{C-D} vibrational bands -- the aromatic \ce{C-D} stretch at 4.40 $\mu$m and the antisymmetric/symmetric aliphatic \ce{C-D} stretch at 4.63/4.75 $\mu$m, respectively \citep{Hudgins2004,Buragohain2015,Buragohain2016}. These bands are detected in regions where the elemental D abundance is typically lower than expected \citep{Peeters2004,Draine2006,Linsky2006,Onaka2013} which suggests that PAHs could be acting as a sink for deuterium. However, there is insufficient data on both the observational and experimental side for a reliable analysis of the amount of deuterium contained in gas-phase PAHs. Large observational studies are hindered by telluric absorptions in the 4--5 $\mu$m range and by the relative weakness of \ce{C-D} bands \citep{Peeters2004,Onaka2013,Doney2016}. Experimental studies are limited to perdeuterated PAHs, known as PADs \citep{Bauschlicher1997,Hudgins1994,Piest2001}. Several processes have been suggested to play a role in interstellar D enrichment of PAHs, varying from gas--grain reactions to gas-phase photodissociation \citep{Sandford2001}. Studies on the contributions and role of these mechanisms have largely focused on solid-state processes, that is, on ices and grains. 

We report on gas-phase unimolecular photodissociation as a possible driver of interstellar D-enrichment of PAHs. We present the photofragmentation mass spectra for UV photolysis of D-enriched protonated anthracene and phenanthrene. We also present their IR-induced fragmentation mass spectra and infrared spectra using infrared multiple-photon dissociation (IRMPD) spectroscopy. We put forward a possible photolysis-induced mechanism and suggest its role in D-enrichment. 

Furthermore, we discuss the astronomical implications of the found mechanism on observations of band intensity ratios for aliphatic and aromatic \ce{C-H}/\ce{C-D} stretch vibrations in PAHs. 

\section{Methods}
\subsection{Experimental methods}
The UV photodissociation mass spectra of protonated (\ce{[H-C14H10]+}), deuteronated (\ce{[D-C14H10]+}), and protonated, perdeuterated (\ce{[H-C14D10]+}) anthracene and phenanthrene were recorded using a Fourier Transform Ion Cyclotron Resonance mass spectrometer (FTICR MS) coupled to a Nd:YAG laser. We applied IRMPD spectroscopy using the Free-Electron Laser for Infrared eXperiments (FELIX) also coupled to the FTICR MS. \citep{Oepts1995,Valle2005}.  Both fragmentation mass spectra and infrared spectral signatures are
provided by IRMPD. This allows for the determination of the molecular structure of the precursor ions. Moreover, as the fragmentation energies of protonated PAHs are significantly lower than those of PAH radical cations, the IR fragmentation mass spectra of protonated PAHs -- unlike their UV counterparts -- are exempt from background signal originating from the isobaric $^{13}$C radical cation isotopolog. This isotopolog is present in a natural abundance of 15.3\%, and cannot be selectively removed according to its mass because of limitations in the mass resolution of our FTICR MS (the $^{12}$CH vs. $^{13}$C mass difference is 0.0045 amu). 

Anthracene and phenanthrene (Sigma Aldrich Co. LLC.; purity > 98\%) were dissolved in methanol at 1 mM concentration, and brought into the gas-phase via electro-spray ionization (ESI) in a Micromass/Waters Z-spray source. Instead of the typically used ammonium acetate or acetic acid, we used {(D-)}trifluoroacetic acid, an efficient protonation agent. The advantage of using D-trifluoroacetic acid in deuterated methanol (\ce{CH3OD} or \ce{CD3OD}) is that $^{1}$H contamination is prevented in the deuteronation process; we experimentally found that such contamination does take place with the use of a weak acid such as ammonium acetate (\citet{Knorke2009}, in water).

Photolysis experiments on the anthracene and phenanthrene ions were performed in the following sequence. The electrosprayed ions were accumulated in a radio-frequency (RF) hexapole trap. They were then pulse-extracted and transported into the ICR cell via a quadrupole bender and a 1m  RF octopole ion guide. The ICR cell was at room temperature and at a pressure of approximately $10^{-8}$ mbar. The precursor ions were mass-isolated by expelling unwanted masses using a Stored Waveform Inverse Fourier Transform (SWIFT) pulse \citep{Marshall1985}. These ions  were then either irradiated in one pass of the UV laser, or with multiple passes of the IR laser beam in a multi-pass configuration, after which all precursor and fragment ions were mass-analyzed. This sequence was repeated three to five times for each wavelength step with a storage and irradiation time of between 2 and 8 s. 

For the UV photolysis experiment, the fourth harmonic of the Nd:YAG laser at 266 nm (4.6611 eV/photon) was used at a pulse energy of $\sim$1 mJ, operated at 10 Hz. For the IRMPD experiment, FELIX was operated at a repetition rate of 10 Hz using macropulses of 5 $\mu$s. The frequency range covered 700--1800 cm$^{-1}$ ($\sim$14--5.5 $\mu$m). The energy per macropulse had a maximum of roughly 65 mJ, and decreased to around 20 mJ at the 1800 cm$^{-1}$(5.5 $\mu$m) edge of the spectral range studied. The spectral bandwidth (FWHM) of FELIX was set to $\sim$0.5\% of the central frequency, which translates into 5 cm$^{-1}$ at 1000 cm$^{-1}$ (10 $\mu$m). A grating spectrometer with an accuracy of $\pm$ 0.01 $\mu$m was used to calibrate the laser wavelength. The laser frequency was changed in steps of 5 cm$^{-1}$. For measurements focusing on the weaker IR modes, the ions were additionally irradiated for 40 ms with the output of a 30 W cw \ce{CO2} laser directly after each FELIX pulse in order to enhance the on-resonance dissociation yield \citep{Settle1992, Almasian2012}. 

\subsection{Theoretical methods}
Density functional theory (DFT) calculations presented in this work were all performed using the Gaussian 09 package \citep{g09}. The B3LYP functional \citep{becke, lee} was used with the 6-311++G(2d,p) basis set to optimize the molecular structure and evaluate the vibrational spectra of the molecules within the harmonic oscillator approximation. B3LYP has been shown previously to accurately predict the vibrational spectra of PAHs \citep{bauschlicher2010}. A scaling factor of 0.966 was applied to the calculated frequencies, allowing the harmonic calculations to line up with the intrinsically anharmonic measured spectra. The theoretical stick spectra were convoluted with a Gaussian line shape with a FWHM of 45 cm$^{-1}$ for anthracene and 60 cm$^{-1}$ for phenanthrene to match the experimental spectrum. As explained by \citet{Oomens2006}, DFT calculations assume a linear absorption process, whilst the IRMPD process is inherently nonlinear. Band positions are most often well reproduced, but larger deviations may be observed for the band intensities.

The potential energy surface of the molecules was investigated using the Minnesota functional M06-2X with the same basis set and was corrected for zero-point vibrational energies. This choice was motivated by the improved accuracy of M06-2X in predicting barrier energies and relative energies of isomers with respect to B3LYP \citep{Zhao2008}. Transition states (TSs) connecting the different structures of the
\ce{[H-C14D10]+} isomers -- in both anthracene and phenanthrene -- were
found with the Berny algorithm. To check that the calculated transition states connected the considered minimum structures, we animated the only mode with an imaginary frequency to confirm the proposed mechanism. Visualization of structures and vibrational modes was performed using the open-source program Gabedit \citep{gabedit}.

\section{Results and Discussion}

\subsection{Anthracene}
\subsubsection{UV and IR photofragmentation mass spectra}
\begin{figure*}
  \centering
    \includegraphics[width=0.9\linewidth]{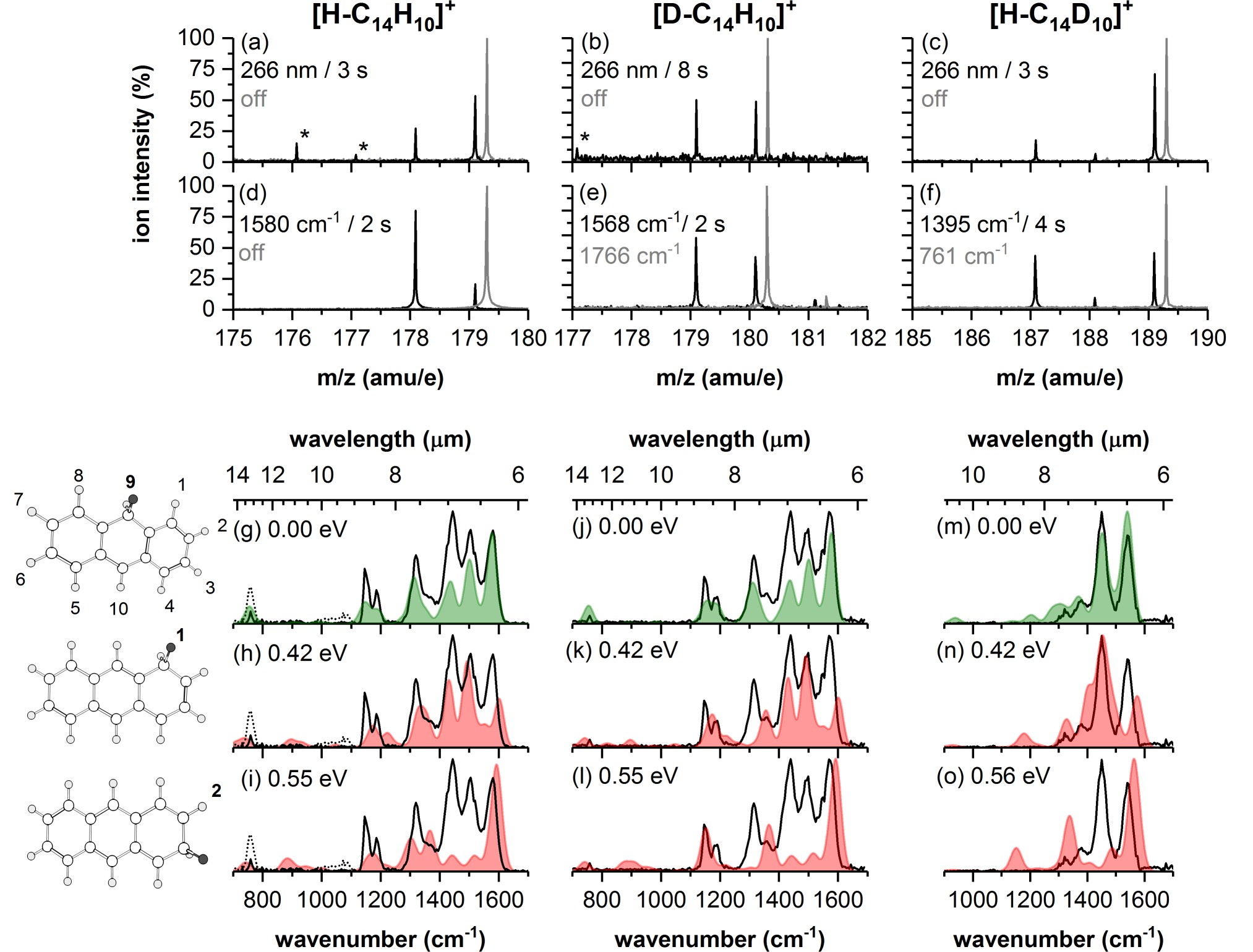}
      \caption{Fragmentation mass (panels a-f) and IRMPD spectra (g-o) recorded for
      three different anthracene isotopologs: protonated anthracene
      -- \ce{[H-C14H10]+}, m/z=179.09 (left column); deuteronated
      anthracene -- \ce{[D-C14H10]+}, m/z=180.09 (middle column);
      protonated, perdeuterated anthracene -- \ce{[H-C14D10]+},
      m/z=189.09 (right column). For all mass spectra, black traces
      represent the fragmentation mass spectra, while gray traces are
      reference precursor mass spectra (with laser on or off-resonance). The
      gray traces are shifted up 0.2 amu to enhance their visibility.
      The top row are mass spectra following UV irradiation, the
      second row following IR irradiation at the indicated IR
      frequency.  Experimental IR spectra are shown in black,
      superimposed on color shaded calculated spectra for each species
      with the protonation (deuteronation) site indicated in the
      structures left of the panels. The theoretical IR spectra were calculated using B3LYP/6-311++G(2d,p) and the ground-state energies in M06-2X/6-311++G(2d,p). Spectra that are considered to ``match'' are colored green; the ones that are considered to not match are red. Relative energies for each
      structure are shown above their respective spectra. The dotted
      curve in panels (g,h,i) represents the measurement assisted by
      the \ce{CO2}-laser. }
      \label{ant_total}
\end{figure*}

We present both the UV and IR photodissociation mass spectra measured for the three isotopologs of anthracene. Figure \ref{ant_total} shows the mass spectra obtained with UV (a,b,c) and IRMPD photodissociation (d,e,f). The gray curves show the precursor mass spectra recorded without irradiation, or with the IR laser off-resonance -- each offset by 0.2 amu for visualization. The black curves depict the fragmentation mass spectra after irradiation. 

Figure \ref{ant_total} (a) shows the UV measurement for protonated anthracene \ce{[H-C14H10]+}, m/z $=$ 179.09 amu. After irradiation, the precursor peak is depleted by approximately 50\%. Additionally, three fragment peaks are observed, which correspond to one to three H atom losses. The loss of one H atom leads to the formation of the radical cation (m/z = 178.08 amu). The two other fragment peaks denoted by asterisks most likely come from the dissociation of the radical cation and its \ce{^{13}C} isotopolog respectively -- which both primarily lose two H atoms -- as reported previously by \citet{Ekern1998} and confirmed by our UV measurements (see Fig. \ref{fig:cation}). 

Analogous measurements for deuteronated anthracene \ce{[D-C14H10]+} are shown in Fig. \ref{ant_total} (b), m/z $=$ 180.09 amu. Upon UV irradiation, the precursor mass is again depleted by approximately 50\%, and one fragment peak is observed that corresponds to the neutral loss of 1 amu. There is a slight indication (asterisk) of subsequent fragmentation of the radical cation as observed in Fig. \ref{ant_total} (a). Therefore, the deuteronated anthracene ion, like the protonated anthracene ion, only loses single hydrogen atoms.

Finally, Fig. \ref{ant_total} (c) shows the UV measurement for protonated,perdeuterated anthracene \ce{[H-C14D10]+}, m/z $=$ 189.14 amu. The precursor mass is slightly depleted in the fragmentation mass spectrum, and two fragment mass peaks are observed, associated with the loss of 1 and 2 amu, the former being the loss of an H atom and the latter the loss of a D atom. Importantly, the loss of 2 amu must correspond to the loss of a D atom as the loss of two H atoms is not possible here. The H/D loss ratio based on the integrated fragment intensities is 28\%/72\%, with an uncertainty of $\pm$5\%.

The IRMPD mass spectrum of protonated anthracene \ce{[H-C14H10]+} is depicted in Fig. \ref{ant_total} (d). A largely depleted precursor peak can be seen and, as expected, no dissociation of the radical cation is observed. Similar to the UV measurements, we only find the loss of single H atoms. For deuteronated anthracene \ce{[D-C14H10]+} (Fig. \ref{ant_total} (e)) the fragmentation mass spectrum again shows a largely depleted precursor and a high-intensity fragment peak, corresponding to single H atom loss, similar to our observations for Fig. \ref{ant_total} (b). Lastly, in Fig. \ref{ant_total} (f), the fragmentation mass spectrum for protonated, perdeuterated anthracene \ce{[H-C14D10]+} shows a precursor peak depleted by approximately half, and two fragment masses corresponding to the loss of H and D, similar to that observed in Fig. \ref{ant_total} (c). The H/D loss ratio based on the integrated intensities is lower at 14\%/86\%, with an uncertainty of $\pm$1\%.

\subsubsection{Infrared spectra}
The IRMPD measurements also yield vibrational spectra that can be used to identify the structure of the precursor ion. Panels (g-o) of Fig. \ref{ant_total}  present the IRMPD spectra of all three anthracene isotopologs (black curves), and compare them to our DFT calculated spectra for the different possible position isomers (shaded traces). For a detailed comparison, the reader is referred to the Appendix, where experimental and theoretical intensities and line positions are given in Tables \ref{table1}-\ref{table3}.

The IR spectra of protonated anthracene \ce{[H-C14H10]+} are shown in panels (g,h,i). The experimental spectrum displays at least seven features dominated by a triad of bands in the 1400-1600 cm$^{-1}$ spectral range that are associated with \ce{C-C} stretching vibrations. In the 1100-1400 cm$^{-1}$ range, three \ce{C-H} in-plane bending modes are observed. The best agreement is found between these peak positions and those of the predicted spectrum for the 9-isomer (g), corresponding to the lowest-energy isomer for \ce{[H-C14H10]+}. The predicted spectra for the other isomers (at 0.42 and 0.55 eV higher in energy) agree less well: mismatches are observed for the experimental band at 1581 cm$^{-1}$ and no bands are observed between 800 and 1000 cm$^{-1}$, where  bands are predicted for both other isomers. To ensure there are no bands in this frequency range, an additional measurement using a \ce{CO2} laser was performed to increase the intensity of the weak features (dotted curve). This only led to an intensity enhancement of the already observed band at 759 cm$^{-1}$, which is a \ce{C-H} out-of-plane quarto (i.e. involving four H atoms) bending vibration, but not to other bands, supporting the assignment of the 9-isomer. This result agrees with those of  \citet{Knorke2009}, who earlier reported an IRMPD spectrum of protonated anthracene in the 1000-1800 cm$^{-1}$ range.

The IR spectra of deuteronated anthracene \ce{[D-C14H10]+} are given in Fig. \ref{ant_total} (j,k,l). The shape is very similar to the spectrum of protonated anthracene, from which we can conclude that exchanging one hydrogen atom for a deuterium has little effect on the IR spectrum in the studied range. The largest difference is that the shoulder of the 1315 cm$^{-1}$ band is now more clearly resolved, allowing us to identify the feature at 1360 cm$^{-1}$ as a separate band associated with an in-plane \ce{C-H} bending vibration. The calculated spectrum for the lowest-energy 9-isomer again agrees best with the experimental spectrum. 

The spectrum of protonated, perdeuterated anthracene \ce{[H-C14D10]+} is displayed in the right column (Fig. \ref{ant_total} (m,n,o)). The perdeuteration simplifies the spectrum, with only four clearly observed bands. Due to the higher mass of D, the \ce{C-D} out-of-plane bending modes shift to lower frequencies, leaving only the \ce{C-C} stretching vibrations in the 6--8 $\mu$m range. The shape of the spectrum with two dominant peaks is excellently predicted by the calculated spectrum of the 9-isomer, depicted by the green curve in Fig. \ref{ant_total} (m). Panels (n) and (o) show that the 1- and 2-isomers provide a significantly poorer match than the 9-isomer, making its assignment to the 9-isomer facile.

\subsubsection{Discussion}
The UV photofragmentation of fully aromatic radical PAH ions leads to predominantly sequential loss of two H atoms and to a lesser extent to loss of \ce{H2}  \citep{Ekern1998, Ling1998, RodriguezCastillo2018}. In both cases, the result is a loss of two H-atoms. The loss of  single H or D atoms must therefore be attributed to a loss from protonated/deuteronated species, which possess aromatic \ce{C-H}/\ce{C-D} sites and one aliphatic \ce{C-H}/\ce{C-D} site. Our calculations show a binding energy for H-loss (D-loss) from an aliphatic site of 2.6~eV (2.7~eV), whereas from an aromatic site this energy  is 4.7~eV (4.8~eV), in agreement with a recent experimental study \citep{West2018}. 

Therefore, H/D loss is very likely from aliphatic sites only. Both isotopologs only have a single aliphatic site from which H or D can be eliminated. The zero-point-energy-corrected binding energy for aliphatic \ce{C-D} is only marginally higher than for \ce{C-H}, so large differences in the H/D loss rates are \textit{a priori} not expected. 
We observe that for \ce{[D-C14H10]+}, photofragmentation only results in the loss of single H atoms, for both IRMPD and UV dissociation. For \ce{[H-C14D10]+}, the opposite appears to happen, with mainly single D atom loss, and a small H atom loss channel. 
The observed large deviations from a 50/50 H/D loss ratio suggest that the molecule undergoes dynamics before dissociation, where the extra H or D atom attached in the protonation/deuteronation process has migrated away from the aliphatic site.

\begin{figure}
  \centering
    \includegraphics[width=\linewidth]{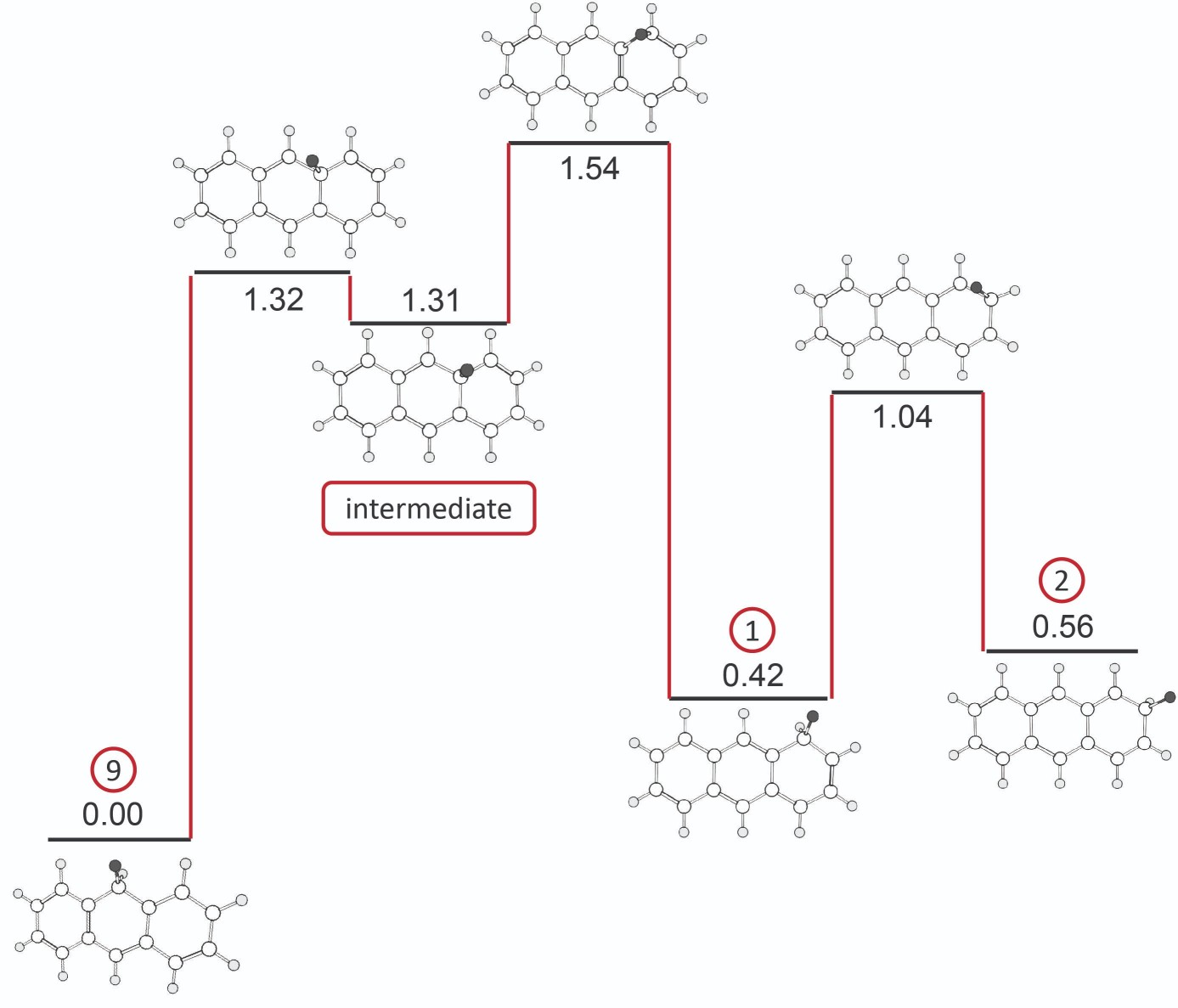}
      \caption{Potential energy surface for D migration linking
      the isomers of protonated, perdeuterated anthracene, (\ce{[H-C14D10]+}) with
      energies in eV, calculated using M06-2X/6-311++G(2d,p). The migrating D atom is highlighted in
      black.}
      \label{ant_TS}
\end{figure}

To investigate this migration further, we consider a mechanism in which H and D atoms can move from one carbon site to the next, leading to a population of \ce{[D-C14H10]+} with a mix of aliphatic \ce{C-HH} and \ce{C-HD} groups, or in \ce{[H-C14D10]+} with a mix of aliphatic \ce{C-DD} and \ce{C-HD} groups. Taking the 1-isomer of \ce{[D-C14H10]+} as an example, the H atom of the \ce{C-HD} group can migrate to the neighboring 2-site, turning it into an aliphatic \ce{C-HH} group while leaving an aromatic \ce{C-D} site behind. This will lead to an increased H/D loss ratio in the photodissociation mass spectra. Alternatively, the D atom can move, resulting in a shift of the \ce{C-HD} on site 1 to a \ce{C-HD} on site 2. This shift will however not have consequences for the fragmentation propensities.

This 1,2-hydrogen (or deuterium) shift is a well-known phenomenon in organic chemistry \citep{Whitmore1932, Kuck2002, Bruice2014}, and specifically for reactions in aromatic molecules \citep{Brooks1999}. Furthermore, in theoretical studies of PAH species, the 1,2-H shift across the PAH rim has been shown to occur once they are excited to internal energies above 1 eV, leading to the formation of various intermediate isomers containing a \ce{C-HH} group \citep{Jolibois2005, Trinquier2017, Castellanos2018a}. Using DFT, we calculated the potential energy reaction pathway for D migration along \ce{[H-C14D10]+}, and the results are depicted in Fig. \ref{ant_TS}. The molecular structures of the three unique position isomers are shown, connected through transition states, including one relatively stable intermediate state where the D is out-of-plane bound to a tertiary carbon. This reactive pathway reveals that the highest barrier is 1.54 eV, which is well below the \ce{C-D} bond fragmentation energy of 2.7 eV. In the current experiments, the IR spectra indicate the presence of only the lowest energy protonation isomer, but we are unable to discern different aromatic \ce{C-D} attachment sites for the observed 9-isomer (see Fig. \ref{ant_shift_sup}). While we cannot fully rule out that 1,2-H or -D shifts occur during the ESI, the absence of different position isomers that are approximately 0.5 eV higher in energy suggests that such high barriers are not overcome in the ESI. On the other hand, we can be certain they are energetically allowed upon photoexcitation, given the energy requirements for fragmentation.

\begin{table*}
\centering
\caption{Barrier heights E$_{b}$ calculated for 1-to-2 shift reaction for the studied species, and RRKM calculated reaction rates at selected internal energies E$_{i}$.}

\label{rates}
\begin{tabular}{lllllll}
        & \multicolumn{1}{l}{} & \multicolumn{1}{l|}{}                                 & \multicolumn{3}{l}{RRKM rates (s$^{-1}$)} \\ 
        & \multicolumn{1}{l}{} & \multicolumn{1}{l|}{}                                 & \multicolumn{3}{l}{for E$_{i}$ (eV)}  \\ \hline
Species & \multicolumn{1}{l}{Atom} & \multicolumn{1}{l|}{E$_{b}$ (eV)}               & 0.64        & 2.7   & 4.7    \\ \hline
\ce{[D-C14H10]+}    & \multicolumn{1}{l}{H} & \multicolumn{1}{l|}{0.61}   & 4.1$\cdot10^{3}$         & 3.2$\cdot10^{9}$      & 3.3$\cdot10^{10}$      \\
\ce{[D-C14H10]+}    & \multicolumn{1}{l}{D} & \multicolumn{1}{l|}{0.64}   & 4.1$\cdot10^{2}$         & 2.3$\cdot10^{9}$      & 2.5$\cdot10^{10}$     \\ \hline
\ce{[H-C14D10]+}    & \multicolumn{1}{l}{H} & \multicolumn{1}{l|}{0.61}   & 1.3$\cdot10^{3}$         & 2.0$\cdot10^{9}$      & 2.4$\cdot10^{10}$     \\
\ce{[H-C14D10]+}    & \multicolumn{1}{l}{D} & \multicolumn{1}{l|}{0.64}   & 58        & 1.2$\cdot10^{9}$      & 1.6$\cdot10^{10}$       
\end{tabular}
\end{table*}

\begin{figure}
\centering
\includegraphics[width=0.8\linewidth]{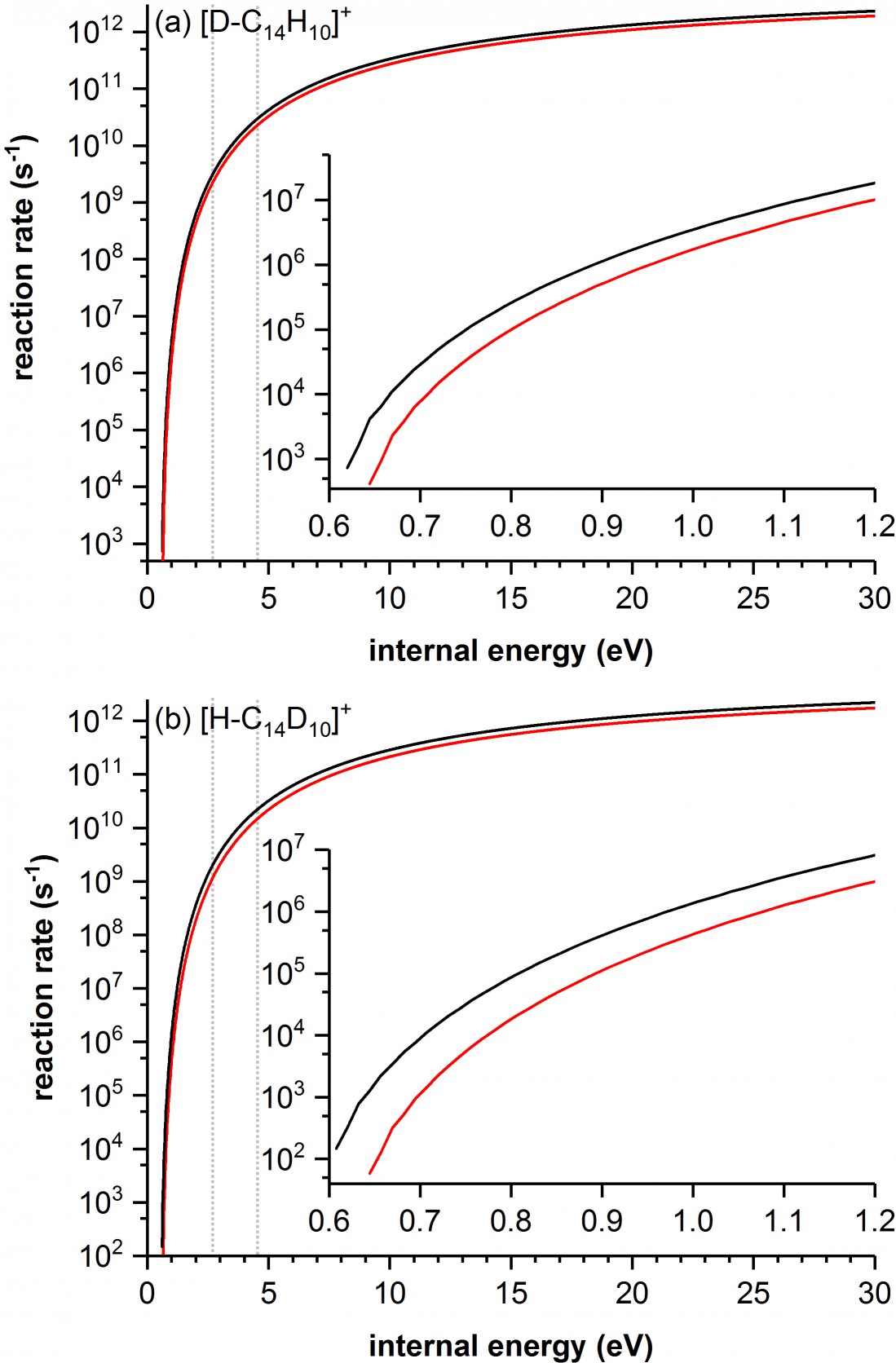}
\caption{RRKM reaction rate calculations for the 1-to-2 H/D-shifts of (a) deuteronated anthracene, \ce{[D-C14H10]+} and (b) protonated, perdeuterated anthracene, \ce{[H-C14D10]+} from the 1-isomer to the 2-isomer, plotted as a function of the internal energy of the molecule in eV. In both (a) and (b), the black curves depict the hydrogen shifting rates and the red curves depict the D shifting rates. Dotted lines depict internal energies attained through IR (min. 2.7 eV) and UV (max. 4.7 eV) excitation}
\label{RRKM}
\end{figure}

In order to assess whether the mobility of atomic H or D is sufficiently high to rationalize the observed propensities for H- and D-loss for \ce{[D-C14H10]+} and \ce{[H-C14D10]+}, it is important to get quantitative information on the different reaction rates involved. We performed RRKM theory rate calculations (\citet{Baer1997}; see \citet{Castellanos2018a} for a detailed description) for both H shifts and D shifts on both isomers, starting from the 1-isomer and moving towards the 2-isomer (see Fig. \ref{ant_TS} for a visual representation). The 1-to-2 shift was chosen as an example due to its relative simplicity compared to the 1-to-9 shift, which would involve two transition states and one intermediate state. The rates plotted as a function of internal energy can be found in Fig. \ref{RRKM} and the rates for three specific energies are listed in Table \ref{rates}. These three energies represent the highest  calculated energy barrier for H/D-shifts (0.64 eV in \ce{[D-C14H10]+}, see Table \ref{rates}), the lowest energy required for H-loss (2.7 eV), and the UV photon energy (4.7 eV), respectively. We observe the following: 1) The H-shift rate is consistently higher than that of the D shift. 2) The lower the internal energy, the larger the difference in H- and D-shift rates. For 0.64 eV, the reaction rates differ by an order of magnitude. 3) The shift rates for both the \ce{[D-C14H10]+} and \ce{[H-C14D10]+} isomers plateau around $10^{12}$ s$^{-1}$ with increasing internal energies. However, shift rates on \ce{[H-C14D10]+} are lower and exhibit a more pronounced difference between the H- and D-shifting rates than for \ce{[D-C14H10]+}.  4) Finally, and most importantly, all rates are so high at the internal energies achieved in both IR and UV experiments that a reaction equilibrium would be reached well within the time frame of our experiment. Although the barriers for other shift reactions are higher (e.g., from the 1- to the 9-position), the current rates are at least four orders of magnitude larger than our experimental time window (10$^{-5}$ s). It is therefore clear that shifts to all different sites on the molecule will occur. This is the equilibrium state in which a mixture of different position isomers exists that will differ between isotopologs, and that we refer to as full scrambling.

\begin{table}
\centering
\caption{Observed relative propensities or H- and D-loss for IR and
    UV experiments.}
\label{loss}
\begin{tabular}{l|cc|cc|cc}
\hline\hline
         & \multicolumn{2}{c|}{UV excited} & \multicolumn{2}{c|}{IR excited} & \multicolumn{2}{c}{Full scrambling}  \\ \hline
         & H               & D             & H               & D  & H               & D                           \\ \hline
\ce{[D-C14H10]+} & 100             & 0             & 100             & 0   & 95&5                        \\
\ce{[H-C14D10]+} & 28              & 72            & 14              & 86   & 5&95                    
\end{tabular}
\end{table}

Purely statistically speaking and assuming full scrambling, 10\% of the aliphatic groups in both the \ce{[D-C14H10]+} and \ce{[H-C14D10]+} isotopologs are \ce{C-HD} groups. Disregarding any difference in probability to eliminate an H or D atom from that group, one would expect 5\% D-loss from \ce{[D-C14H10]+} and 5\% H-loss from \ce{[H-C14D10]+}. Table \ref{loss} lists these statistical factors together with the measured UV and IR photofragmentation factors of both isotopologs. Comparison with the experiment shows that the measured hydrogen loss exceeds the statistical probability, whereas the measured deuterium loss falls short in both isotopologs. This difference can be attributed to the differences in the migration rates discussed above, which leads to different mobility  and loss rates for the D and H atoms.

From these migration rates, it can be seen that H is more likely to shift than  D, which in \ce{[D-C14H10]+} results in an isomer in which an aromatic \ce{C-D} is left behind and an aliphatic \ce{C-HH} is created. From the perspective of this new \ce{C-HH}, the next shift will either lead to a new \ce{C-HH} or back to the \ce{C-HD}. If the D had begun to shift, the \ce{C-HD} site would have shifted position, and the odds would be higher for the next shift to be with the H atom, both leading to a \ce{C-HH}. This means that not only does the zero-point energy difference in binding energy lead to preferential H atom loss in the case of a \ce{C-HD} site, but also that the larger mobility will lead to preferential creation of \ce{C-HH} sites. In the case of \ce{[H-C14D10]+}, the H atom will simply create another \ce{C-HD} site. The shift of a D atom will lead to a \ce{C-DD} site from which D atom loss may occur.

\subsection{Phenanthrene}
We  further investigated the scrambling mechanism by performing experiments on phenanthrene-D$_{10}$, a three-ringed PAH with a non-linear structure.

\subsubsection{Infrared photofragmentation mass spectrum}
\begin{figure}
    \centering
    \caption{Infrared photofragmentation (panel a, black trace) and
      reference precursor (gray trace, shifted up 0.2 amu) mass spectra, and IRMPD spectrum (panels
      (b-f), black trace) for protonated, perdeuterated phenanthrene --
      \ce{[H-C14D10]+}.  Panels (b--f) further contain calculated spectra (blue) for
      five different position isomers, for which the structures are shown on the left. The theoretical IR spectra were calculated using B3LYP/6-311++G(2d,p) and the ground-state energies in M06-2X/6-311++G(2d,p).}
    \label{phen-all}
    \includegraphics[width=\linewidth]{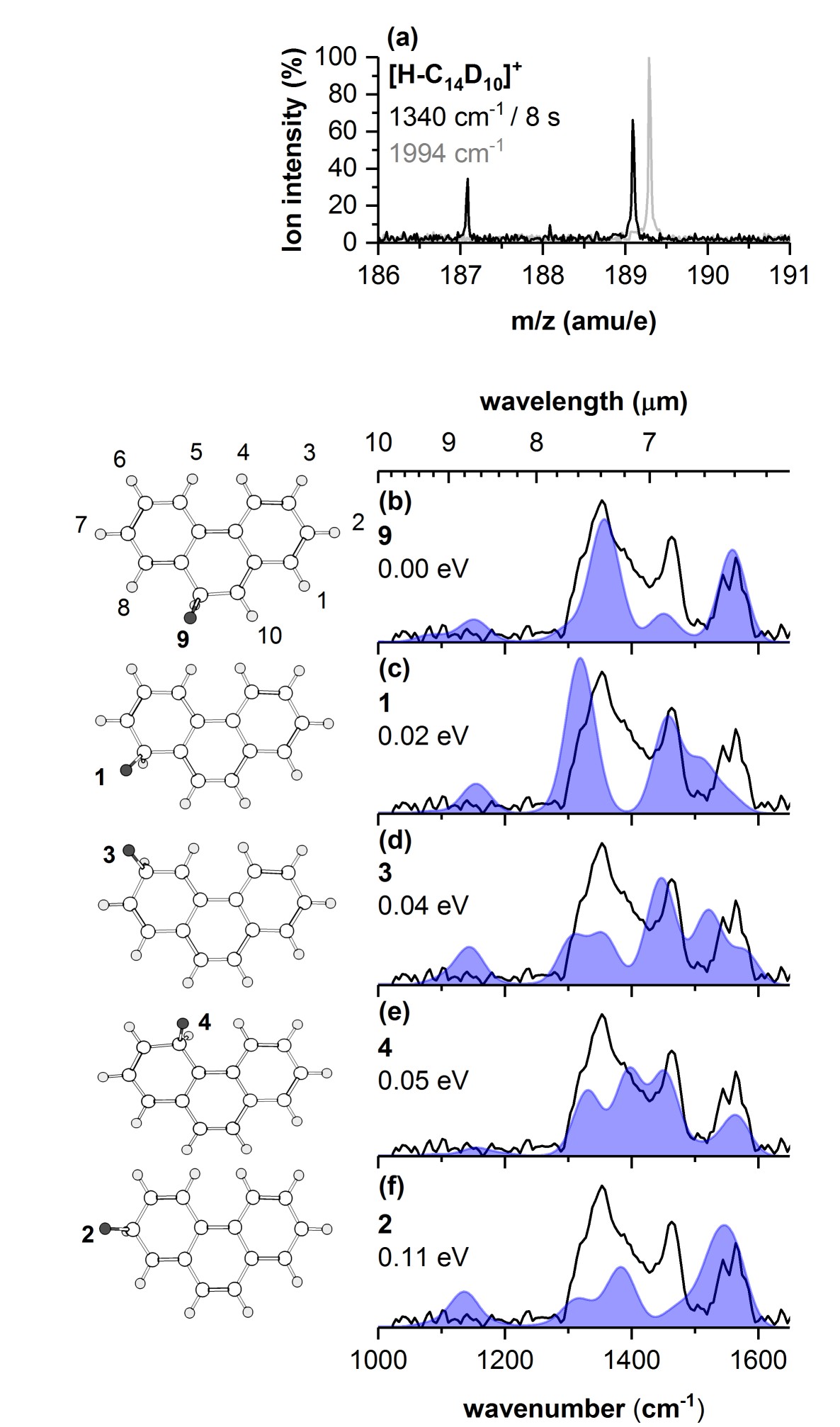}
\end{figure}

Figure \ref{phen-all} (a) shows the IRMPD mass spectrum of protonated, perdeuterated phenanthrene (\ce{[H-C14D10]+}), m/z $=$ 189.14 amu. The color coding, offsetting, and normalization are identical to that of the anthracene mass spectra. In the fragmentation mass spectrum, the precursor peak is depleted by more than 30\% and two fragment peaks are observed, displaying the dominant loss of 2 amu and smaller loss of 1 amu, amounting to an integrated H/D loss ratio of 14\%/86\% with an uncertainty of $\pm$ 4 \%.

\subsubsection{Infrared spectrum}
The experimental IR spectrum of \ce{[H-C14D10]+} is displayed by the black curves in Fig.\ref{phen-all} (b--f). The theoretical spectra for the five possible structural isomers are shown as the blue shaded traces. They are listed in order of increasing energy, from top to bottom. The experimental IR spectrum shows three main features in the \ce{C-C} stretching region between 1300 and 1600 cm$^{-1}$. No features appear in the 1000-1200 cm$^{-1}$ region. The shape is roughly reproduced by the predicted lowest energy 9-isomer in Fig. \ref{phen-all} (b). The 4-isomer (Fig. \ref{phen-all} (e)), which lies only 0.05 eV higher, also matches the experiment well. The remaining three theoretical spectra do not reproduce the shape of the features in the experiment, and show significantly more IR activity in the 1000-1200 cm$^{-1}$ region than the 9-isomer and 4-isomer. We nevertheless conclude that it is likely that under our experimental conditions several isomers are present in significant amounts, in contrast to anthracene where only one isomer was present. Although this hinders a conclusive assignment of the IR spectrum, it does not affect the discussion on the proposed scrambling mechanism as becomes clear below. 

\subsubsection{Discussion}
\begin{figure*}
  \centering
    \includegraphics[width=0.6\linewidth]{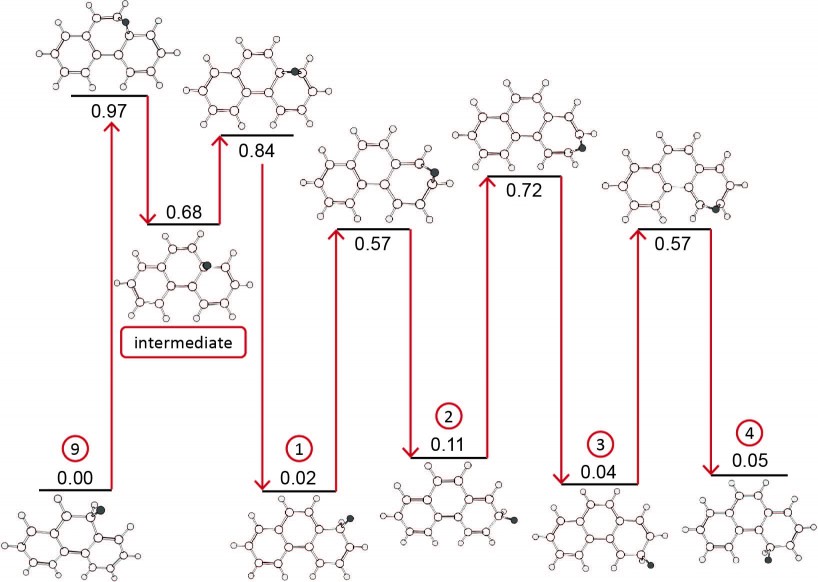}
      \caption{Isomers of protonated, perdeuterated phenanthrene and their transition states, with corresponding energies in eV calculated with M06-2X/6-311++G(2d,p).}
      \label{phen_TS}
\end{figure*}

We find a clear indication that phenanthrene-\ce{D10} undergoes a scrambling mechanism similar to that for anthracene-\ce{D10}. As can be observed in Fig. \ref{phen-all} (a), phenanthrene-\ce{D10} exhibits dominant D atom loss and little H atom loss, a behavior very similar to anthracene-\ce{D10} in Fig. \ref{ant_total} (c,f). The H/D loss ratio of phenanthrene is the same as the ratio for anthracene, albeit with a larger uncertainty. It is interesting that the measured H/D ratios of IRMPD-fragmented \ce{[H-C14D10]+} are the same for both anthracene and phenanthrene, considering that their potential energy surfaces are different. The protonation isomer energies of phenanthrene are an order of magnitude closer together than those of anthracene. At room temperature, it is likely that the precursor phenanthrene ions exist in a mixture of multiple isomers, whereas only one precursor 9-isomer is present for anthracene. 

The structural difference leads to differing energy barriers and therefore scrambling rates (see Figs. \ref{ant_TS} and \ref{phen_TS}). The major rate-limiting step for full scrambling in phenanthrene as well as in anthracene is the shift from the 9- to the 1- position, for which the hydrogen has to move across a nonhydrogenated tertiary carbon, which is bound to three other carbons. Crossing this tertiary carbon is associated with an energy barrier of 1.54 eV for anthracene, which is 1.6 times higher than the analogous barrier of 0.97 eV for phenanthrene. Scrambling is therefore expected to be substantially faster for phenanthrene. Moreover, the initial aliphatic \ce{C-HD} group in anthracene is located on a solo site from which migration to either neighboring sites is associated with a barrier of 1.54 eV. Thus, for any -- even partial -- scrambling to occur before photolysis of anthracene, this highest barrier needs to be overcome. For phenanthrene, the initial aliphatic \ce{C-HD} group is located on a duo or quarto site,  already allowing for partial scrambling to occur on the same aromatic ring at relatively low energies. Partial scrambling on nonsolo sites is therefore not only faster, but also possible at lower excitation energies or temperatures. That the measured H/D ratios are the same for both \ce{[H-C14D10]+} isomers indicates that full scrambling is easily achieved before dissociation for these small molecules. This makes sense considering that our experiment takes place on microsecond timescales, and our calculated rates imply a much faster scrambling process. 

\section{Astrophysical implications}
Gas-phase hydrogenation of PAH molecules has been studied both theoretically and experimentally \citep[see][]{thrower2012,klerke2013,cazaux2016,Vala2017,Cazaux2019,Ferullo2019}, because it plays a role in shaping the PAH populations in the more shielded environments of photodissociation regions \citep[PDRs; ][]{Montillaud2013,boschman2015,Andrews2016}. Singly and even multiply hydrogenated PAHs are able to maintain their hydrogenation state in these shielded regions, allowing for the catalytic formation of \ce{H2} \citep{Wakelam2017}. For large (>50 carbons) molecules, multi-photon excitation is required for H-loss to occur \citep{Montillaud2013,Andrews2016}. The mechanism for these interstellar loss processes has been studied experimentally, and the roaming of hydrogen atoms on hydrogenated PAHs has been recently put forward to explain the experimental results for H/\ce{H2} photodissociation in large PAHs \citep{Castellanos2018a, Castellanos2018b}. 

\citet{Doney2016} detected deuterium-containing PAHs via their \ce{C-D} stretches in astronomical objects with intense aliphatic \ce{C-H} bands. In these regions, the aliphatic to aromatic H ratio inferred from the ratio of the 3.4 and 3.29 $\mu$m band intensities is 0.2-0.3, almost an order of magnitude larger than typically observed for PAHs in the ISM \citep{Tielens2008}. This suggests that D-containing PAHs are predominantly present in regions that favor hydrogenation or deuteration. These regions correspond to the more shielded layers of PDRs, with $G_0/n(H)$ values smaller than 0.03 \citep{Andrews2016}. The additional H/D atom preferably attaches to the edge of hydrogenated PAH molecules, with low to no barriers, creating aliphatic groups \citep{rauls2008}. Our study shows that upon hydrogenation or deuteration, and irradiation, PAHs are prone to facile scrambling.

The calculated reaction barriers for the 1,2-H/D shifts (shown in Figs. \ref{ant_TS} and \ref{phen_TS}) are at least 1.2 eV lower than the energies supplied by the photo-excitation processes in this study, and are clearly below the energy of interstellar UV radiation \citep{Allamandola1989,Montillaud2013}. This implies that for PAHs that are able to withstand interstellar radiation without losing H or D atoms, scrambling would inevitably occur. As shown by the RRKM rate calculations, 1,2-H shifts have higher reaction rates than the equivalent 1,2-D shifts, which creates a bias towards aliphatic \ce{C-HH} groups over \ce{C-HD} groups on PAHs with low deuteration levels (1 or 2 D atoms). While the D-PAH molecule is exposed to a stronger UV field, aliphatic H will be preferably lost over D. Similarly, this mechanism could lead to the uptake and preservation of multiple D atoms on the molecule, up to a point where the higher probability of \ce{C-DD} site formation leads to D loss being  favored over H loss. From a spectroscopic point of view, this process could be traced looking at the intensity variation of the aliphatic \ce{C-H}/aromatic \ce{C-D} stretching bands (3.4/4.4 and 3.5/4.4 $\mu$m) with a spatially resolved study of an extended source like IRAS12073-6233, which shows hints of spatial variation of the D/H ratio \citep{Doney2016}. 

Large, compact PAHs -- which are believed to be the most abundant class of PAHs in the ISM \citep{Ricca2012} -- are expected to exhibit more pronounced effects. Solo sites are more abundant on compact PAHs and are highly reactive \citep{Aihara1996}, meaning that D is most likely to attach itself there and stay there due to the meta-stable tertiary carbon neighbor sites (see Figs. \ref{ant_TS} and \ref{phen_TS}). Furthermore, the absorbed photon energy will be distributed over many more vibrational modes in a large PAH than in small PAHs, leading to lower average excitation per vibrational state. Because the difference in scrambling and dissociation behavior of H and D atoms is larger at lower energies, this means that the larger the PAH molecule, the more pronounced the mobility difference between D and H. Based on these considerations, we speculate that large, compact PAHs with an aliphatic \ce{C-HD} group on solo sites might cause a shift in the position of the aliphatic \ce{C-D} stretching bands around 4.75 $\mu$m, as was also observed in calculations for deuteronated ovalene by \citet{Buragohain2016}. A full calculation for large PAHs is beyond the scope of this paper, and will be included in future work \citep{Wiersma2020}. The upcoming \textit{James Webb Space Telescope} mission will provide improved spatial resolution and sensitivity compared to Spitzer and AKARI, which should make it possible to observe new regions and provide us with greater detail for the already well-known ones. Finally, new photochemical models including the effect of scrambling are needed to determine the role of PAHs as possible deuterium reservoirs.

\section{Conclusion}
We  present the UV and IR photofragmentation mass spectra of deuterium-containing isotopologs of protonated anthracene and phenanthrene. These fragmentation mass spectra demonstrate that singly deuteronated molecules do not show D atom loss, while the protonated, perdeuterated molecules show dominant D atom loss and low H atom loss. Supported by DFT and RRKM rate calculations of 1,2-H/D shift reactions, we show that these observations can only be explained by a mechanism that allows H and D to roam over the peripheral carbon atoms, leading to a ``scrambling'' of different isomers, which will readily occur upon absorption of interstellar UV photons. The scrambling of the H and D atoms across the carbon skeleton rim has several important implications for the chemistry of larger PAHs in the ISM. One of these implications is that large PAHs can take up several deuterium atoms by ``locking'' them on aromatic sites. This would suggest that the deuterium fractionation on PAHs could be higher than previously modeled. We also postulate that aliphatic \ce{C-HD} solo sites may be more abundant than other configurations, which is due to the elevated barriers for 1,2-H/D shifts across carbon sites and the relative reactivity of solo sites. To test our hypotheses, new photochemical modeling and a re-evaluation of observational data are needed.

\subsection*{Acknowledgements}
We gratefully acknowledge the {\it Nederlandse Organisatie voor Wetenschappelijk Onderzoek} (NWO) for the support of the FELIX Laboratory. This work is supported by the VIDI grant (723.014.007) of A.P. from NWO. Furthermore, A.C. gratefully acknowledges NWO for a VENI grant (639.041.543). Calculations were carried out on the Dutch national e-infrastructure (Cartesius and LISA) with the support of Surfsara, under projects NWO Rekentijd 16260 and 17603. Part of this work was inspired by the COST Action CM1401 'Our Astro-Chemical History'. S.W. would also like to thank dr. Mridusmita Buragohain for the fruitful discussions.

% for the bibliography, at the end
\bibliographystyle{aa} % style aa.bst
\bibliography{PhotolysisofPAHs} % your references Yourfile.bib

\begin{appendix}

\section{Anthracene} 

\subsection{Anthracene cation mass spectrum}

\begin{figure}
  \centering
    \includegraphics[width=0.65\linewidth]{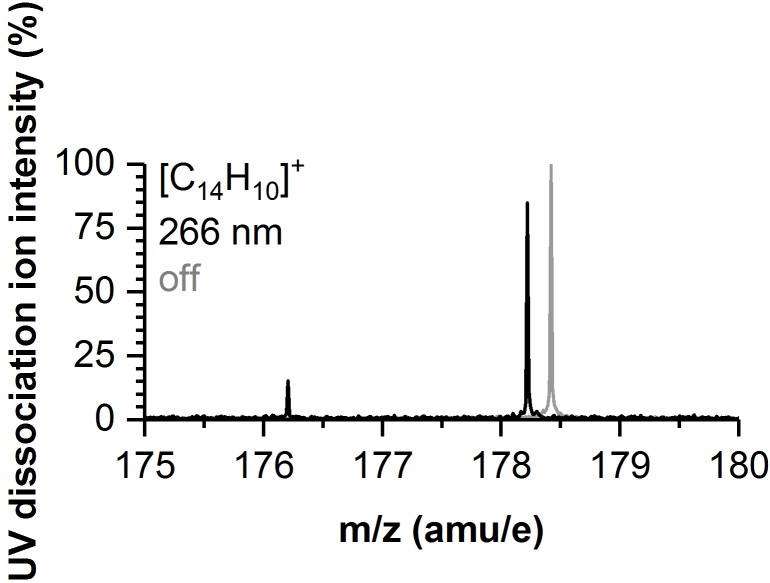}
      \caption{Fragmentation mass spectrum of the anthracene cation \ce{C14H10+}. The black trace was recorded after exposure to 1 mJ pulses at 10 Hz of a 266 nm (4th harmonic of a Nd:YAG laser) for 3 s, while the gray trace was recorded with the laser off and was given an offset of 0.2 amu for legibility.}
      \label{fig:cation}
\end{figure}
Figure \ref{fig:cation} shows fragmentation mass spectra for the radical cation \ce{C14H10+} at m/z $=$ 178 amu (gray). After UV irradiation (black), the fragmentation mass spectrum shows one fragment at m/z $=$ 176 amu, depicting the loss of two H atoms as the sole dissociation process from the radical cation, which was also observed by \citet{Ekern1998}. 

A quantitative interpretation of mass spectra exhibiting sequential H loss could be complicated by the presence of naturally abundant \ce{^{13}C} isotopologs of the PAH under investigation, amounting to 15.3\% for a molecule with 14 carbons. To ensure the isotopic purity, the mass spectra were made as clean as possible prior to isolation, that is, minimizing the intensity of the cations in the case of protonation, and optimization of the deuteronation versus the protonation for the deuteronated anthracene. Using unisolated mass spectra, it is possible to make accurate estimates of the isotopic contamination in the precursor peaks for each spectrum. Knowing that the cation only loses two hydrogen atoms, we can discern contributions of the protonated and deuteronated isomers from the contributions of the radical cationic \ce{^{13}C} isomers. 

\subsection{Anthracene band positions}

\begin{table*}[ht]
\centering

\caption{Band positions in cm$^{-1}$ of the experimental spectrum of protonated anthracene \ce{[H-C14H10]+} and the most intense modes of the unconvoluted, theoretical spectra, capped off at 10 km/mol. A scaling factor of 0.9662 has been applied to the calculated spectrum to correct for anharmonicity. The experimental IR intensities are normalized to the highest intensity, while the theoretical IR intensities are the absolute cross sections in km/mol.}
\begin{tabular}{ll||ll|ll|ll}
\multicolumn{8}{c}{\ce{[H-C14H10]+}}\\
\hline\hline
\multicolumn{2}{l||}{experiment} & \multicolumn{2}{l|}{9-isomer} & \multicolumn{2}{l|}{1-isomer} & \multicolumn{2}{l}{2-isomer} \\
\hline
$\bar{\nu}$           & I        & $\bar{\nu}$       & I       & $\bar{\nu}$       & I         & $\bar{\nu}$         & I      \\
(cm$^{-1}$)           & (a.u.)          & (cm$^{-1}$)         & (km/mol)         &  (cm$^{-1}$)          & (km/mol)          & (cm$^{-1}$)         & (km/mol)                 \\
\hline
759        & 0.11       & 755          & 84            & 743          & 41            & 741          & 38            \\
           &            &              &               &              &               & 873          & 34            \\
           &            &              &               &              &               & 886          & 11            \\
           &            &              &               &              &               & 888          & 16            \\
           &            &              &               &              &               & 900          & 15            \\
           &            &              &               & 932          & 25            & 911          & 11            \\
           &            &              &               &              &               & 944          & 20            \\
           &            & 984          & 10            & 1047         & 15            & 1005         & 11            \\
1146       & 0.49       & 1146         & 88            & 1163         & 25            & 1155         & 52            \\
           &            & 1166         & 15            &              &               &              &               \\
1185       & 0.30       & 1191         & 63            & 1177         & 88            & 1181         & 68            \\
           &            &              &               & 1223         & 70            & 1221         & 28            \\
           &            & 1276         & 20            & 1257         & 12            & 1261         & 13            \\
           &            &              &               &              &               & 1290         & 74            \\
1319       & 0.61       & 1312         & 220           & 1324         & 139           & 1309         & 113           \\
           &            & 1337         & 18            &              &               &              &               \\
           &            & 1347         & 14            & 1335         & 35            & 1333         & 26            \\
           &            & 1355         & 43            & 1355         & 128           & 1364         & 163           \\
           &            &              &               & 1376         & 22            & 1381         & 49            \\
           &            & 1409         & 65            & 1406         & 13            & 1419         & 12            \\
           &            & 1438         & 168           & 1432         & 324           &              &               \\
1445       & 1.00       & 1441         & 25            &              &               & 1441         & 66            \\
           &            & 1476         & 18            & 1492         & 404           & 1466         & 25            \\
1505       & 0.84       & 1501         & 305           &              &               & 1506         & 32            \\
           &            & 1535         & 11            & 1521         & 85            & 1523         & 56            \\
           &            & 1539         & 35            & 1555         & 101           & 1568         & 99            \\
1581       & 0.83       & 1578         & 429           & 1590         & 14            & 1592         & 443           \\
           &            & 1592         & 21            & 1602         & 228           & 1611         & 103          
\end{tabular}
\label{table1}
\end{table*}

\begin{table*}[]
\centering

\caption{Band positions in cm$^{-1}$ of the experimental spectrum of deuteronated anthracene \ce{[D-C14H10]+} and the most intense modes of the unconvoluted, theoretical spectra, capped off at 10 km/mol. A scaling factor of 0.9662 has been applied to the calculated spectrum to correct for anharmonicity. The experimental IR intensities are normalized to the highest intensity, while the theoretical IR intensities are the absolute cross sections in km/mol.}
\begin{tabular}{ll||ll|ll|ll}
\multicolumn{8}{c}{\ce{[D-C14H10]+} }                                                                                           \\
\hline\hline
\multicolumn{2}{l||}{experiment} & \multicolumn{2}{l|}{9-isomer} & \multicolumn{2}{l|}{1-isomer} & \multicolumn{2}{l}{2-isomer} \\
\hline
$\bar{\nu}$           & I        & $\bar{\nu}$       & I       & $\bar{\nu}$       & I         & $\bar{\nu}$         & I      \\
(cm$^{-1}$)           & (a.u.)          & (cm$^{-1}$)         & (km/mol)         &  (cm$^{-1}$)          & (km/mol)          & (cm$^{-1}$)         & (km/mol)                 \\
\hline
758        & 0.07       & 755          & 86            & 743          & 44            & 737          & 13            \\
           &            &              &               &              &               & 742          & 26            \\
           &            &              &               & 811          & 14            & 869          & 29            \\
           &            &              &               & 828          & 10            & 888          & 20            \\
           &            &              &               & 896          & 35            & 913          & 28            \\
           &            & 984          & 10            & 1048         & 17            & 950          & 14            \\
1146       & 0.42       & 1150         & 81            & 1159         & 50            & 1142         & 33            \\
           &            & 1158         & 12            & 1177         & 95            & 1152         & 137           \\
           &            & 1167         & 19            &              &               & 1155         & 46            \\
1191       & 0.24       & 1188         & 66            & 1183         & 37            & 1183         & 26            \\
           &            & 1201         & 25            & 1223         & 54            & 1220         & 29            \\
           &            & 1276         & 20            & 1265         & 10            & 1261         & 12            \\
1315       & 0.61       & 1308         & 178           & 1319         & 19            & 1307         & 15            \\
           &            & 1333         & 54            &              &               &              &               \\
1360       & 0.27       & 1337         & 16            & 1355         & 168           & 1363         & 194           \\
           &            & 1406         & 55            & 1374         & 25            & 1380         & 48            \\
           &            &              &               & 1403         & 13            &              &               \\
           &            &              &               & 1424         & 6             &              &               \\
1439       & 1.00       & 1438         & 180           & 1432         & 334           & 1441         & 65            \\
           &            & 1441         & 24            &              &               & 1466         & 27            \\
           &            & 1474         & 25            &              &               &              &               \\
1499       & 0.85       & 1501         & 305           & 1490         & 434           & 1505         & 42            \\
           &            & 1534         & 13            & 1520         & 81            & 1523         & 61            \\
           &            & 1538         & 35            & 1554         & 92            & 1567         & 93            \\
1569       & 1.00       & 1577         & 429           & 1590         & 10            & 1592         & 473           \\
           &            & 1592         & 21            & 1602         & 237           & 1609         & 95           
\end{tabular}
\label{table2}
\end{table*}

\begin{table*}[]
\centering

\caption{Band positions in cm$^{-1}$ of the experimental spectrum of protonated, perdeuterated anthracene \ce{[H-C14D10]+} and the most intense modes of the unconvoluted, theoretical spectra, capped off at 10 km/mol. A scaling factor of 0.9662 has been applied to the calculated spectrum to correct for anharmonicity. The experimental IR intensities are normalized to the highest intensity, while the theoretical IR intensities are the absolute cross sections in km/mol.}
\begin{tabular}{ll||ll|ll|ll}
\multicolumn{8}{c}{\ce{[H-C14D10]+}}                                                                                        \\
\hline\hline
\multicolumn{2}{l||}{experiment} & \multicolumn{2}{l|}{9-isomer} & \multicolumn{2}{l|}{1-isomer} & \multicolumn{2}{l}{2-isomer} \\
\hline
$\bar{\nu}$           & I        & $\bar{\nu}$       & I       & $\bar{\nu}$       & I         & $\bar{\nu}$         & I      \\
(cm$^{-1}$)           & (a.u.)          & (cm$^{-1}$)         & (km/mol)         &  (cm$^{-1}$)          & (km/mol)          & (cm$^{-1}$)         & (km/mol)                 \\
\hline
           &            & 938          & 29            & 932          & 11            &              &               \\
           &            & 1139         & 13            &              &               & 1152         & 134           \\
           &            & 1197         & 13            & 1179         & 74            &              &               \\
           &            & 1206         & 31            &              &               &              &               \\
           &            &              &               & 1227         & 14            & 1233         & 14            \\
           &            & 1277         & 67            &              &               & 1289         & 24            \\
           &            & 1311         & 70            &              &               &              &               \\
1321       & 0.13       & 1324         & 13            & 1327         & 112           & 1337         & 291           \\
           &            & 1349         & 28            & 1334         & 39            & 1349         & 33            \\
1380       & 0.21       & 1371         & 110           & 1402         & 263           & 1391         & 9             \\
           &            & 1389         & 19            & 1403         & 46            & 1410         & 43            \\
1450       & 1          & 1451         & 451           & 1452         & 552           & 1473         & 16            \\
           &            & 1494         & 73            & 1486         & 169           & 1488         & 120           \\
           &            & 1499         & 27            & 1506         & 74            & 1532         & 11            \\
1540       & 0.79       & 1540         & 555           &              &               & 1560         & 524           \\
           &            & 1559         & 14            & 1575         & 273           & 1577         & 162          
\end{tabular}
\label{table3}
\end{table*}
Table \ref{table1} lists all of the measured band positions of protonated anthracene featured in Fig. \ref{ant_total} (g,h,i), and compares them with the theoretical band positions of the three different position isomers as calculated using DFT. 

Table \ref{table2} lists those measured and calculated for deuteronated anthracene given in Fig. \ref{ant_total} (j,k,l). Interestingly, the only band not predicted by the 9-isomer is precisely the experimental band at 1360 cm$^{-1}$. Suitable modes are predicted at 1355 cm$^{-1}$ for the 1-isomer (k) and at 1363 for the 2-isomer (l). However, the predicted features of the 1- and 2-isomers around 900 cm$^{-1}$ are not present in the experiment. Overall, the features are consistent with the 9-isomer.

Table \ref{table3} lists all measured and calculated for protonated, perdeuterated anthracene given in Fig. \ref{ant_total} (m,n,o). Only calculated frequencies that exhibit enough intensity to allow for comparison with the experimental features are listed.

\subsection{Theoretical spectra for additional anthracene isomers}
\begin{figure*}
  \centering
    \includegraphics[width=0.65\linewidth]{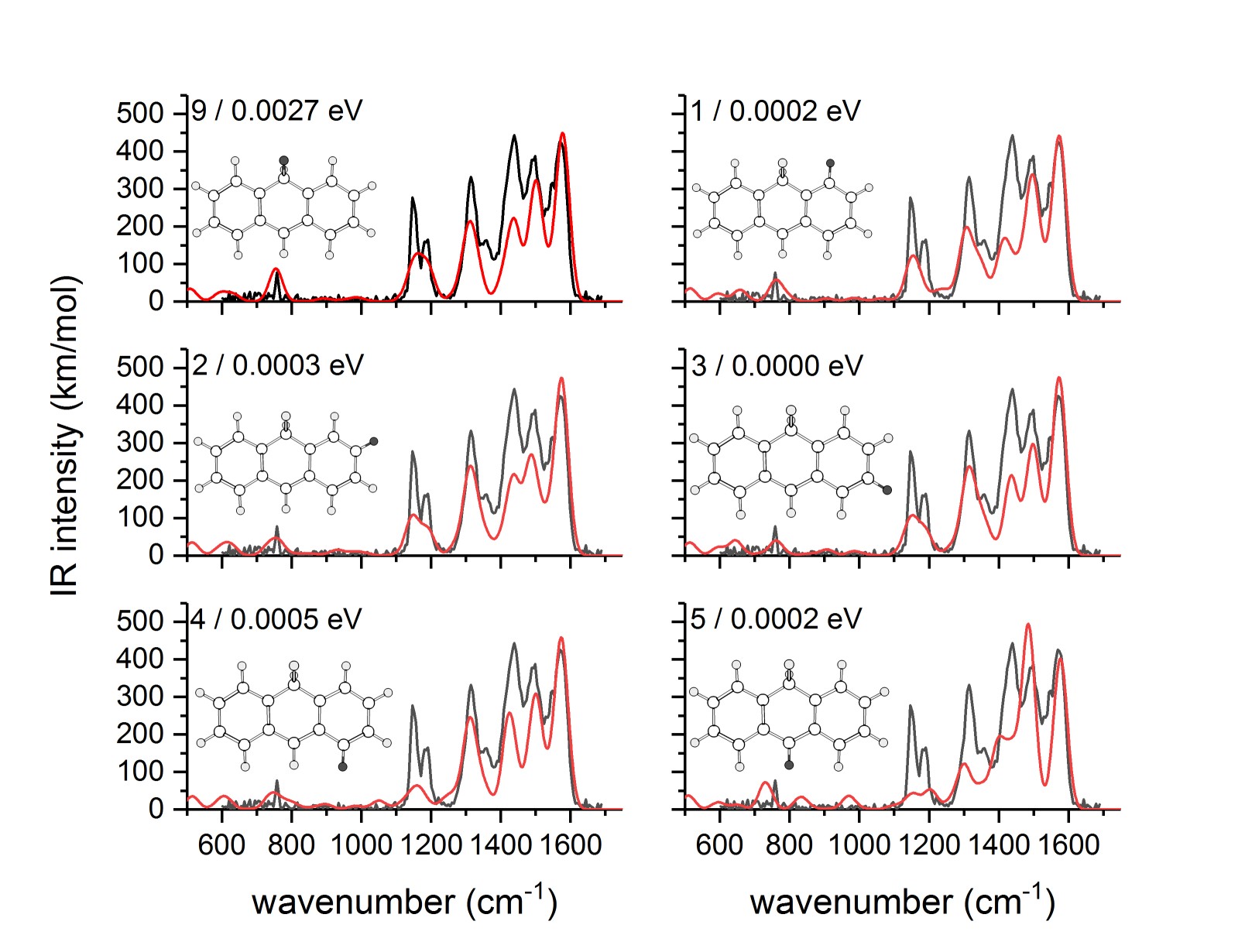}
      \caption{Calculations for the six scrambling isomers for the 9-isomer of deuteronated anthracene (red) compared to the FELIX IRMPD spectrum of \ce{[D-C14H10]+} (black). The shifting D atom is marked in black on the molecule, and the label in the upper right corner of each panel indicates its position. The energies of each of these isomers is listed in eV with respect to the global minimum. }
      \label{ant_shift_sup}
\end{figure*}

Figure \ref{ant_shift_sup} displays the six different isotopic isomers that can be formed by a \ce{[D-C14H10]+} with the aliphatic group at the 9-position. These spectra mostly differ in terms of intensity ratios. The only spectrum showing significant changes in band position, is the one where the D is at the 5 site, opposite the \ce{C-HH} group, which is the most symmetric configuration. This increased symmetry leads to improved resonant enhancement of the in-plane \ce{C-H} bending vibrations, which significantly changes the shape of the spectrum. However, this theoretical spectrum is clearly not in accord with the experimental spectrum.

\section{Phenanthrene}
\subsection{Band positions for phenanthrene}
\begin{table*}[]
\centering
\caption{Band positions in cm$^{-1}$ of the experimental spectrum of protonated, perdeuterated phenanthrene \ce{[H-C14D10]+} and the most intense modes of the unconvoluted, theoretical spectra, capped off at 10 km/mol. A scaling factor of 0.9662 has been applied to the calculated spectrum to correct for anharmonicity. The experimental IR intensities are normalized to the highest intensity, while the theoretical IR intensities are the absolute cross sections in km/mol.}

\begin{tabular}{ll||ll|ll|ll|ll|ll}
\multicolumn{12}{c}{\ce{H+C14D10}}                                                                                                                                                             \\
\hline\hline
\multicolumn{2}{c||}{experiment} & \multicolumn{2}{c|}{9-isomer} & \multicolumn{2}{c|}{1-isomer} & \multicolumn{2}{c|}{3-isomer} & \multicolumn{2}{c|}{4-isomer} & \multicolumn{2}{c}{2-isomer} \\
\hline
$\bar{\nu}$           & I        & $\bar{\nu}$       & I       & $\bar{\nu}$       & I         & $\bar{\nu}$         & I       & $\bar{\nu}$       & I          & $\bar{\nu}$      & I          \\
(cm$^{-1}$)           & (a.u.)          & (cm$^{-1}$)         & (km/mol)         &  (cm$^{-1}$)          & (km/mol)          & (cm$^{-1}$)         & (km/mol)           & (cm$^{-1}$)         & (km/mol)          & (cm$^{-1}$)         &(km/mol)           \\
\hline
               &               & 1080          & 24           &               &              & 1108          & 22           &               &              &               &              \\
               &               & 1119          & 15           &               &              &               &              &               &              &               &              \\
               &               & 1152          & 74           & 1153          & 81           & 1145          & 128          & 1154          & 29           & 1137          & 120          \\
               &               &               &              & 1168          & 23           &               &              &               &              &               &              \\
               &               & 1298          & 44           &               &              & 1303          & 60           &               &              &               &              \\
               &               &               &              & 1314          & 394          & 1308          & 103          &               &              & 1314          & 83           \\
               &               &               &              & 1317          & 22           &               &              & 1323          & 48           &               &              \\
1353           & 1             & 1351          & 284          & 1332          & 165          & 1352          & 125          & 1332          & 187          & 1345          & 31           \\
               &               & 1367          & 169          &               &              & 1365          & 55           &               &              &               &              \\
               &               & 1387          & 23           &               &              &               &              & 1387          & 46           & 1382          & 182          \\
               &               &               &              &               &              &               &              & 1398          & 262          &               &              \\
               &               &               &              &               &              & 1435          & 125          &               &              & 1402          & 39           \\
1463           & 0.75          & 1450          & 99           & 1456          & 323          & 1452          & 277          & 1451          & 284          & 1465          & 48           \\
               &               &               &              &               &              &               &              &               &              & 1496          & 65           \\
               &               &               &              &               &              & 1520          & 238          & 1483          & 24           & 1532          & 232          \\
               &               & 1551          & 44           & 1510          & 170          & 1541          & 29           & 1517          & 29           & 1558          & 100          \\
1565           & 0.59          & 1560          & 287          & 1556          & 54           & 1579          & 115          & 1566          & 134          & 1567          & 140         
\end{tabular}
\label{table4}
\end{table*}
Table \ref{table4} lists all of the measured bands visible in Fig. \ref{phen-all} (b--f), and compares them with the calculated band positions of the five different position isomers.
\end{appendix}

\end{document}